\pdfoutput=1
\documentclass[a4paper,11pt]{article}
\usepackage{pos}

\usepackage{cleveref}
\usepackage{xspace}
\usepackage[cachedir=./,frozencache=true]{minted}
\setminted{fontsize=\small,linenos,numberblanklines=false,framesep=1mm}
\setminted{frame=none}
\usepackage{xcolor}
\usepackage{wrapfig}
\usepackage{siunitx}
\usepackage{subcaption}
\usepackage{epstopdf}

\usepackage{todonotes}

\title{What the new RooFit can do for your analysis}

\author*[a]{Stephan Hageboeck}

\affiliation[a]{CERN,\\
  1, Esplanade des Particules, Meyrin, Switzerland}

\emailAdd{stephan.hageboeck@cern.ch}

\newcommand{\RooFit}{\texttt{RooFit}\xspace}
\newcommand{\RooStats}{\texttt{RooStats}\xspace}
\newcommand{\HistFactory}{\texttt{HistFactory}\xspace}
\newcommand{\ROOT}{\texttt{ROOT}\xspace}
\newcommand{\PDF}{\texttt{PDF}\xspace}
\newcommand{\PDFs}{\texttt{PDF}s\xspace}

\newcommand{\eg}{e.g.\ }

\abstract{
\RooFit is a toolkit for statistical modelling and fitting, and together with \RooStats it is used for measurements and statistical tests by most experiments in particle physics.
Since one year, \RooFit is being modernised.
In this talk, improvements already released with ROOT will be discussed, such as faster data loading, vectorised computations and more standard-like interfaces.
These allow for speeding up unbinned fits by several factors, and make \RooFit easier to use from both C++ and Python.
}

\FullConference{%
  40th International Conference on High Energy physics - ICHEP2020\\
  July 28 - August 6, 2020\\
  Prague, Czech Republic (virtual meeting)
}

\begin{document}
\maketitle

\section{Introduction}
\RooFit~\cite{RooFit} is a C++ package for statistical modelling distributed with \ROOT~\cite{ROOT}.
With \RooFit, users can define likelihood models using observables, parameters, functions and \PDFs \footnote{\texttt{P}robability \texttt{D}ensity \texttt{F}unctions}, which can be fit to data, plotted or be used for statistical tests.
The tools for performing such tests are provided by the \RooStats package, and the \HistFactory package provides tools to create \RooFit models from collections of \ROOT histograms.

\RooFit was started in the year 2000 in the BaBar collaboration.
Since then, \RooFit has been a reliable tool for many experiments in high-energy physics at $B$ factories or the Large Hadron Collider.
Due to its long history, it is nonetheless time to modernise and optimise \RooFit for today's hardware, enabling researchers to analyse larger datasets, to devise more elaborate statistical models and to solve challenging research questions.

\section{Improving the Usability of \RooFit}
\subsection{Extending \RooFit with more Stable and Faster Built-in \PDFs\label{sec:newPDFs}}
\begin{wrapfigure}[9]{r}{0.395\textwidth}
\vspace{-1\baselineskip}
\centering
\includegraphics[width=5cm,trim={8mm 0mm 0 14mm},clip=true]{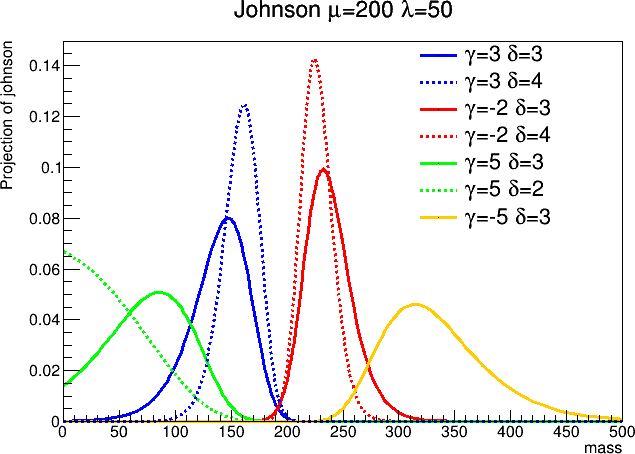}
\vspace{-0.5\baselineskip}
\hfil\caption{Johnson~\cite{Johnson} distribution for various values of the parameters $\gamma$ and $\delta$}
\label{fig:johnson}
\end{wrapfigure}
In modernising \RooFit, the Johnson~\cite{Johnson} (\ROOT 6.18+, \cref{fig:johnson}) and Hypatia2~\cite{Hypatia} distributions (\ROOT 6.20+) were added.
Although \RooFit can interpret any formula using \ROOT's \texttt{cling} interpreter~\cite{cling}, built-in \PDFs are usually more stable, and can be optimised to evaluate faster~\cite{CHEP19}.

In \ROOT 6.20+, another convenience \PDF was added, \texttt{RooWrapperPdf}.
Since \RooFit treats functions and \PDFs differently (the former are just evaluated while the latter are evaluated and normalised automatically), users are sometimes forced to decide whether they should implement an object as a function, as a \PDF or both.
With the addition of \texttt{RooWrapperPdf}, only the function implementation has to be provided.
The function can be used as a \PDF by wrapping it into \texttt{RooWrapperPdf}.
That is, the function is augmented with automatic numerical normalisation, and toy data can be sampled from it.

\subsection{Unbiased Binned Fits}
\begin{figure}
    \centering
    \begin{subfigure}[b]{0.49\textwidth}
        \centering
        \includegraphics[width=\textwidth]{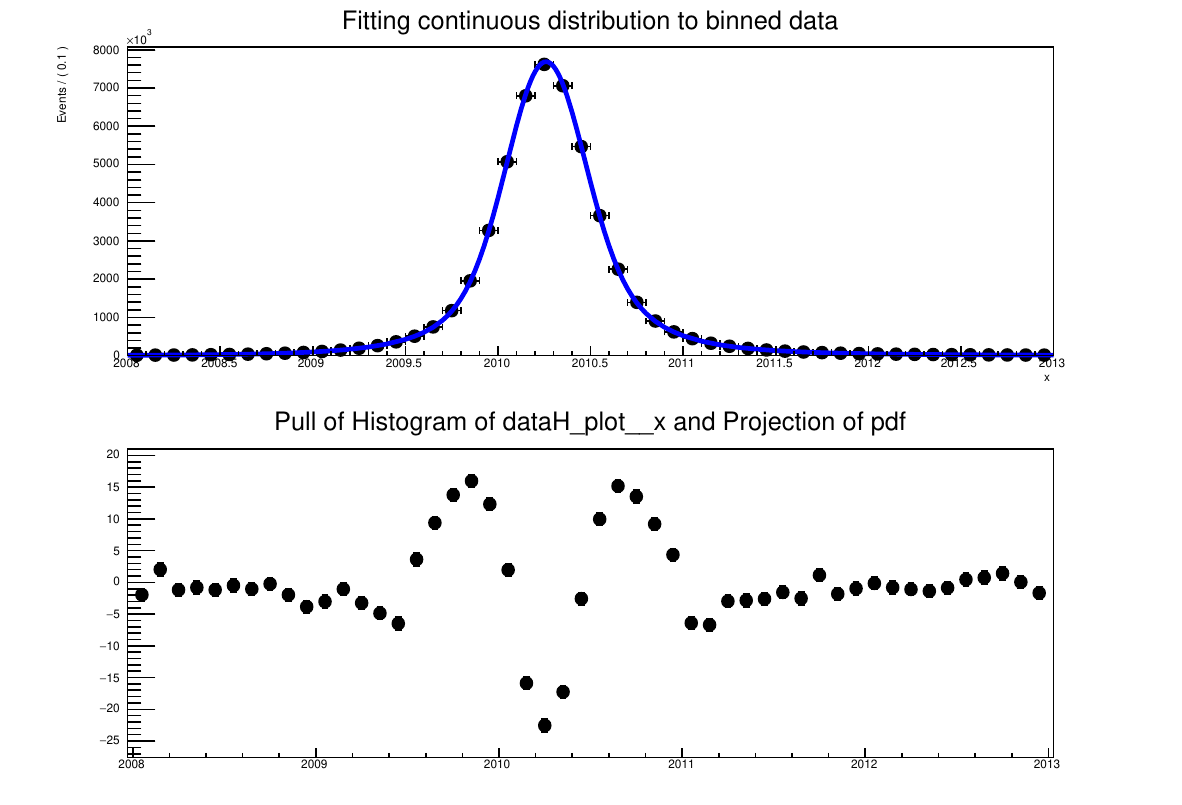}
        \caption{Continuous \PDF compared to binned data}
        \label{fig:fit_continuousPdf}
    \end{subfigure}
    \hfill
    \begin{subfigure}[b]{0.49\textwidth}
        \centering
        \includegraphics[width=\textwidth]{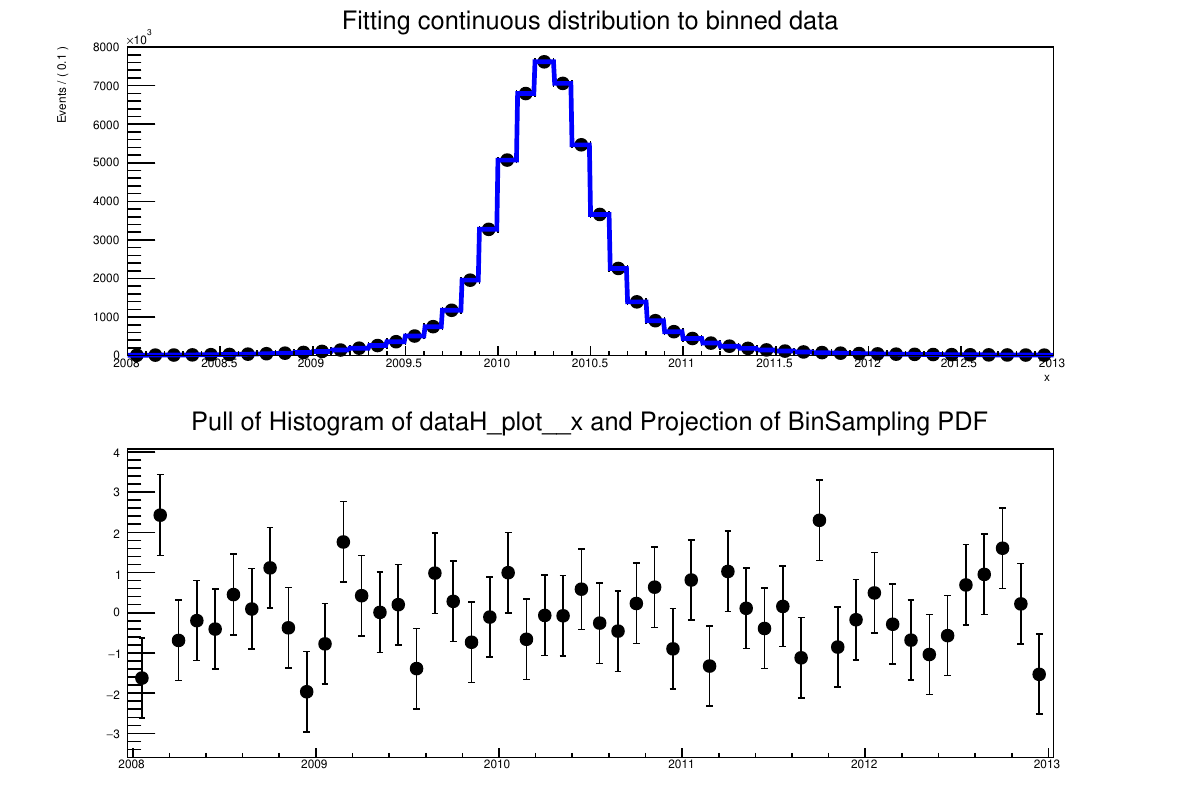}
        \caption{\PDF wrapped into \texttt{RooBinSamplingPdf}}
        \label{fig:fit_binSamplingPdf}
    \end{subfigure}
    \caption{Comparison of pulls with and without \texttt{RooBinSamplingPdf}. The pulls are computed by comparing event counts with the plotted curves at the bin centres. NB: The $y$-axis is zoomed 6-fold in (\subref{fig:fit_binSamplingPdf}).}
    \label{fig:compareBinnedFits}
\end{figure}
A long-known problem in \RooFit discussed during the conference are binned fits with \PDFs that have high curvature\footnote{That is non-zero second derivatives. These are models that cannot easily be approximated with piece-wise linear functions.}.
To save computation time in binned fits, \RooFit evaluates \PDFs only in the centre of a bin, and uses this as an approximation for the probability of an entire bin.
If a function has a high curvature, this is not correct, and can lead to biases as shown in \cref{fig:fit_continuousPdf}.
These biases can be reduced by using more bins, but this is not always an option for users given available data statistics.
In \ROOT 6.24+, the class \texttt{RooBinSamplingPdf} was added to integrate a continuous \PDF over each bin.
It converts continuous into binned \PDFs as shown in \cref{fig:fit_binSamplingPdf}, and evaluating it in the bin centre, in fact, anywhere in a bin, yields correct probability densities.
Residuals are reduced, fits converge more reliably and fit results are not biased, any more.

\subsection{Recovery from Evaluation Errors}
When \RooFit evaluates a mathematical model, it relies on the fact that all model parameters are within the domains of the functions involved.
If this assumption is violated, computations might yield the value \texttt{NaN}\footnote{``Not a Number''. For a Gaussian distribution, for example, this happens when setting the parameter $\sigma \leq 0$.}.
Starting from \ROOT 6.24, \RooFit will warn users if they set up parameters that might violate the domain of a function. Not all invalid parameters can be prevented by checking parameter limits, though.
If the minimiser chooses a set of parameters that violates the domain of a function, the model cannot be evaluated, and no gradient can be computed to continue the minimisation.
The minimiser will try to change each parameter, but it may only slowly, sometimes never, find a set of valid parameters to continue the minimisation.

Starting from \ROOT 6.24, some of these fit failures can be avoided.
\RooFit can pass information to the minimiser by packing it into the mantissa of a \texttt{NaN}.
If, for example, a \PDF evaluates to a negative value, which is disallowed, the magnitude of the undershoot is packed into a \texttt{NaN}.
Since IEEE-compliant floating-point operations leave the mantissa unaffected, this information can propagate through all computations.
Before the log-likelihood is passed to the minimiser, this information is unpacked and the magnitudes of all violations are summed.
This is converted into a penalty term, which is passed to the minimiser.
From the magnitude of the penalty term, the minimiser can compute a gradient, and use it to step away from disallowed parameters.
Notoriously unstable \PDFs such as \texttt{RooPolynomial}\footnote{Since polynomials of uneven order inevitably evaluate to negative values if $x$ is large or small enough, very careful tuning of model parameters is required to keep the polynomial positive across the fit range of the observable.} were found to fit to data much faster, and fits that would previously fail were found to converge more reliably.


\subsection{Modernisation of Interfaces}
\label{sec:c++}
In \RooFit, any collection of mathematical entities such as parameters, observables, functions or \PDFs are saved or passed to functions using the classes \texttt{RooArgSet} and \texttt{RooArgList}.
To operate \RooFit, users have to manipulate these collections, for example, for querying the values of fit parameters.
Until \ROOT 6.18, these collections were based on a linked list shipped with \RooFit. Iterating through such a collection was cumbersome and not efficient.
To allow for the use of range-based \mintinline{c++}{for} loops and to speed up iterations, \RooFit's collections were converted to  \mintinline{c++}{std::vector}-based collections in \ROOT 6.18.
This results in simpler code:\\
\begin{minipage}[t]{0.48\textwidth}
\centering \ROOT 6.18+
\begin{minted}[fontsize=\footnotesize]{c++}
// No variables outside loop required


for (auto p : *pdf.getParameters(obs))
  p->Print();

// No danger of memory leak
\end{minted}
\end{minipage}
\hfil
\begin{minipage}[t]{0.48\textwidth}
\centering \ROOT 6.16 and before\begin{minted}[fontsize=\footnotesize]{c++}
TIterator* it =
    pdf.getParameters(obs)->createIterator();
RooAbsArg* p;
while ((p=(RooAbsArg*)it->Next())) {
  p->Print();
}
delete it;
\end{minted}
\end{minipage}

\smallskip
Iterating proved to be \SI{25}{\%} faster, and the code is significantly simpler.
Typical workflows in \RooFit are sped up from \SIrange{5}{21}{\%}~\cite{ACAT19}, depending on how many iterations through collections are required.
The old interface remains supported, however, so users are not forced to rewrite their code.

Starting from \ROOT 6.22, \RooFit uses modernised category classes, which behave map-like. Defining and printing category states compares as follows:

\begin{minipage}[t]{0.49\textwidth}
\centering \ROOT 6.22+
\begin{minted}[fontsize=\footnotesize]{c++}
RooCategory cat("cat", "Lep. mult.");
cat["0Lep"] = 0;
cat["1Lep"] = 1;

for (const auto& name_idx : cat) {
  std::cout << name_idx.first << ", "
    << name_idx.second << std::endl;
}
\end{minted}
\end{minipage}
\hfil
\begin{minipage}[t]{0.49\textwidth}
\centering \ROOT 6.20
\begin{minted}[fontsize=\footnotesize]{c++}
RooCategory cat("cat", "Lep. mult.");
cat.defineType("0Lep", 0);
cat.defineType("1Lep", 1);

TIterator* typeIt = cat.typeIterator();
RooCatType* catType;
while ( (catType =
  dynamic_cast<RooCatType*>(typeIt->Next()) )
  != nullptr) {
  std::cout << catType.GetName() << ", "
    << catType.getVal() << std::endl;
}
delete typeIt;
\end{minted}
\end{minipage}

\smallskip
The new categories use 4 instead of 288 bytes of memory per entry in a dataset, and can better be integrated into batch computations (see \cref{sec:fastPDFs}).
Also here, old interfaces remain supported.

The modernisation of C++ interfaces is also beneficial for using \RooFit from Python.
Since \ROOT ships with the C++ interpreter \texttt{cling}~\cite{cling}, it can automatically generate Python bindings for C++ objects (``\texttt{PyROOT}''~\cite{Pyroot}).
Before \ROOT 6.18, Python users would have had to imitate the C++ code at the beginning of this section.
Now, the equivalent loop reads:
\begin{minted}{Python}
for p in pdf.getParameters(obs):
  p.Print()
\end{minted}
In addition to the automatically generated interfaces, \texttt{PyROOT} features so-called ``Pythonisations'', short Python code that helps steer C++.
For example, while an import function for \RooFit objects reads
``\mintinline{c++}{workspace.import(object)}'' in C++,
Python users were required to use the workaround
``\mintinline{Python}{getattr(workspace, 'import')(object)}'',
since \texttt{import} is a reserved keyword. Starting from \ROOT 6.22, users can use the more intuitive 
``\mintinline{Python}{workspace.Import(object)}''.

\section{Faster \PDF Computations\label{sec:fastPDFs}}
When \RooFit computes likelihoods, one or multiple \PDFs have to be evaluated for each entry in a dataset.
However, \RooFit only evaluates a single event in each function call\footnote{A multi-process mode can be used to parallelise computations, but this still computes one event per function call.}, and rows of a dataset are read one by one.
This strategy is inefficient, because it does not make use of data caches, memory prefetching or vector extensions.

Starting from \ROOT 6.20, data access and likelihood computations were reorganised such that data are read from a dataset in batches instead of copying them out of a dataset one by one~\cite{CHEP19}.
Each sub-computation produces multiple values per function call, so less functions have to be called.
Data are read in blocks.
This mode uses slightly more memory, because intermediate results have to be stored, but it speeds up likelihood computations by about \SI{300}{\%}, which is especially relevant for data-intensive unbinned fits.
Note that this speed up multiplies with the speed up from multi-processing.

\PDF classes need to implement the interface '\mintinline{c++}{span getValues(...)}'\footnote{In \ROOT 6.20 and 6.22, the experimental \mintinline{c++}{getValBatch()} was in use, but it has been superseded by \mintinline{c++}{getValues()}.} to benefit from this speed up.
Most \PDFs in \RooFit have been updated to support it.
External \PDFs that do not implement this interface can be used nevertheless, since a fall-back function that calls \RooFit's classic evaluation functions in a loop will be used.
Since the optimised functions might yield slightly different numbers, the fast batch mode has to be activated by users:\\
\mintinline[escapeinside=||]{c++}{pdf.fitTo(data, |\textcolor{blue}{BatchMode}|(true)); // Evaluate likelihood using fast batch mode}

\begin{figure}[t]
\begin{center}
\includegraphics[width=1.\linewidth,trim={0 3mm 0 10mm},clip=true]{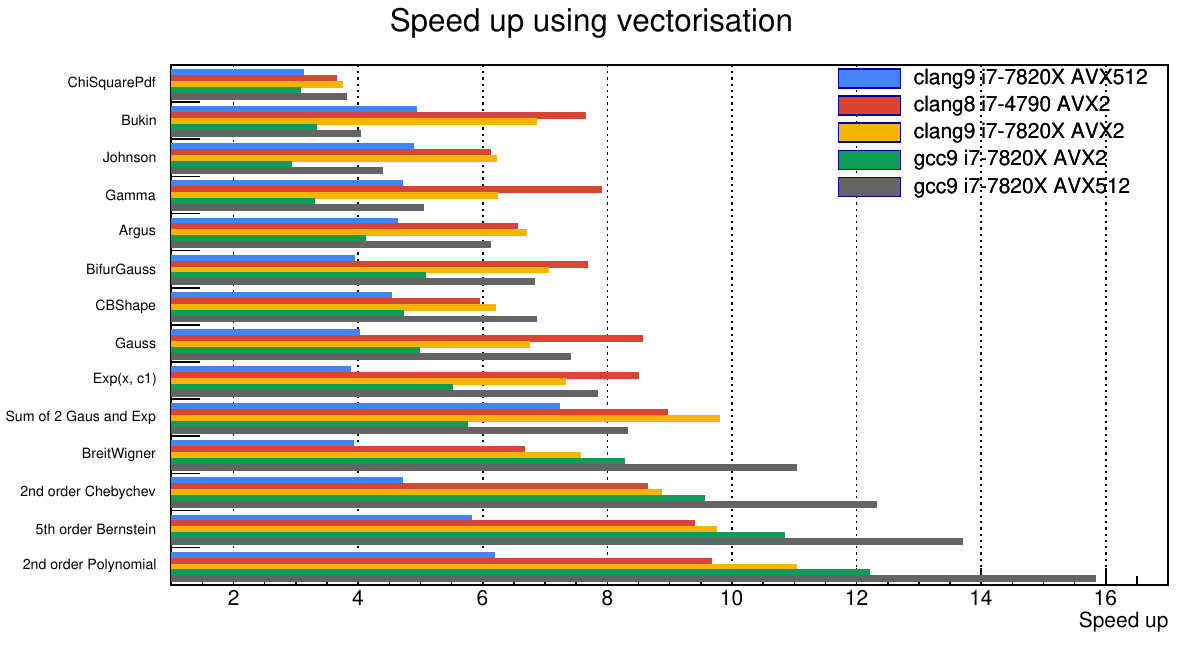}
\end{center}
\caption{Speed up for computing the likelihoods of datasets of $100\,000$ to $300\,000$ events for various likelihood models.
The fast batch interface in \ROOT 6.20 is timed against the normal single-value computations. Depending on compiler, CPU capabilities and workflow, a speed up of 3x to 16x can be expected.}
\label{fig:vecSpeed}
\end{figure}

On modern CPU architectures with vector extensions such as \texttt{SSE} or \texttt{AVX}, the batch computation functions can be optimised even further using \texttt{SIMD} computations~\cite{CHEP19}.
\Cref{fig:vecSpeed} shows the speed up that was achieved in comparison to \RooFit's classic single-value computations when vector extensions are used.
The speed up ranges from 3x for the ``ChiSquared'' \PDF, which is based on a function from an external library without vectorisation, to 16x with \texttt{AVX512} extensions.

To use vector extensions, users have to compile \ROOT 6.20 and 6.22 themselves, manually enabling the desired extensions.
Therefore, pre-compiled \ROOT distributions (\eg provided centrally by collaborations) will mostly benefit from the 3x speed up due to the more efficient data access.
\ROOT 6.24 and later will ship with multiple versions of a \RooFit computation library, which are optimised for different vector extensions.
\ROOT will inspect the CPU and load the fastest library supported by the hardware.
Users will therefore be able to benefit from larger speed ups than 3x.

Unit tests ensure that the optimised computation functions yield the same results as the classic \RooFit functions.
Computations of likelihoods usually agree to a relative accuracy of \num{1.E-14}, and log-likelihoods agree up to \num{2.E-14} with a few exceptions.
Fit parameters usually agree better than \num{1.E-5}, which is orders of magnitude smaller than the statistical error in most fits.

\section{Conclusions}
In 2019, development in \RooFit has been resumed.
\RooFit's interfaces are being modernised, long-standing problems are solved, and significant speed ups were achieved by better using the capabilities of modern CPUs.
Work is underway to use the benefits of the new computation interface for \RooFit computations on GPUs.
These developments are aimed at providing \RooFit's users with a better tool, and the \ROOT team encourages users to get in touch with ideas and requests.

\bibliographystyle{JHEP}
\bibliography{CHEP19}

\providecommand{\href}[2]{#2}\begingroup\raggedright\begin{thebibliography}{1}

\bibitem{RooFit}
W.~Verkerke and D.P.~Kirkby, \emph{{The RooFit toolkit for data modeling}},
  {\emph{econf} {\bfseries C0303241} (2003) MOLT007}
  [\href{https://arxiv.org/abs/physics/0306116}{{\ttfamily physics/0306116}}].

\bibitem{ROOT}
R.~Brun and F.~Rademakers, \emph{{ROOT: An object oriented data analysis
  framework}}, {\emph{Nucl. Instrum. Methods Phys. Res.} {\bfseries A389}
  (1997) 81}.

\bibitem{Johnson}
N.L.~Johnson, \emph{Systems of frequency curves generated by methods of
  translation}, {\emph{Biometrika} {\bfseries 36} (1949) 149}.

\bibitem{Hypatia}
D.M.~Santos and F.~Dupertuis, \emph{Mass distributions marginalized over
  per-event errors},
  \href{https://doi.org/https://doi.org/10.1016/j.nima.2014.06.081}{\emph{Nucl.
  Instrum. Methods Phys. Res. A} {\bfseries 764} (2014) 150 }.

\bibitem{cling}
V.~Vasilev, P.~Canal, A.~Naumann and P.~Russo, \emph{Cling {\textendash} the
  new interactive interpreter for {ROOT} 6},
  \href{https://doi.org/10.1088/1742-6596/396/5/052071}{\emph{J. Phys.: Conf.
  Ser.} {\bfseries 396} (2012) 052071}.

\bibitem{CHEP19}
S.~Hageboeck, \emph{{A Faster, More Intuitive RooFit}},
  \href{https://doi.org/10.1051/epjconf/202024506007}{\emph{EPJ Web Conf.}
  {\bfseries 245} (2020) 06007}
  [\href{https://arxiv.org/abs/2003.12875}{{\ttfamily 2003.12875}}].

\bibitem{ACAT19}
S.~Hageboeck and L.~Moneta, \emph{{Making RooFit Ready for Run 3}},
  \href{https://doi.org/10.1088/1742-6596/1525/1/012114}{\emph{J. Phys. Conf.
  Ser.} {\bfseries 1525} (2020) 012114}
  [\href{https://arxiv.org/abs/2003.12861}{{\ttfamily 2003.12861}}].

\bibitem{Pyroot}
M.~Galli, E.~Tejedor and S.~Wunsch, \emph{{A New PyROOT: Modern, Interoperable
  and More Pythonic}},
  \href{https://doi.org/10.1051/epjconf/202024506004}{\emph{EPJ Web Conf.}
  {\bfseries 245} (2020) 06004}.

\end{thebibliography}\endgroup

\end{document}